# Spatio-Temporal Chaos and Patterns Formation in Nonequilibrium Media: Phenomenological Model of Electronic Turbulence.

E.S.Mchedlova and D.I.Trubetskov
College of Applied Science
Saratov State University,
83 Astrakhanskaya, Saratov 410071,
Russia

The processes in nonequilibrium dissipative media caused by coherent structure formation and lead to the complicated dynamics are of interest for nonlinear physics. Here we consider a model of the flow of interacting electronics patterns. Under electronic pattern we understand a small volume of active media, containing electron oscillators. The superradiation phenomenon takes place due to the evolving radiation instability within each small volume of the flow. It is numerically shown that the dynamics of radiation is changed from spatio-temporal complex oscillations to arising of spatial disorder, i. e. chaotic spatial pattern formation.

## 1. Introduction

Constructing the appropriate phenomenological models is one of the central problems in dissipative nonequilibrium medum investigation. First of all we keep in mind the models that can help to explain the turbulence phenomenon. In particular, the model of spatial chaos development in ensembles of unidirectionally coupled lumped systems are well known for description of hydrodynamical turbulence in the flow of interacting coherent patterns (see, for example, [1]).

One can assume that the turbulence in electronic beams may be explained by the similar models. Really, appearance of patterns in electronic beams was observed experimentally by authors of known paper [2]. In experiment from [3] the chaos was founded in annular electron beam drifting in longitudinal uniform magnetic field. It is the authors' opinion that the complex dynamics of the flow resulted from the pattern interaction.

We shall consider the phenomenological model of electron beam, since the patterns have been devoted.

Before the model will be stated, we have to mention some points on extended system modelling in general. It has been known that there are some mathematical approaches to presentation of extended system. First of them is the most general and corresponds to partial differential equations. After immediate numerical solving them the model takes to some a discreteness. Some other approaches involve the initial approximation of continuous media by some discrete analogy, such as chains and lattices, built up from lumped systems.

Because we are focused on the flow since the patterns have been formed, we dwell on chain of moving interacting elements. The model is of interior complexity: each of its elements (patterns) involves a limited number of subunits - ensembles of nonlinear oscillators. In such case each pattern cannot be assumed to be a lumped system. As for discreteness of the model, it completely agrees with our supposition that the patterns exist in the flow.

The structure of paper is as follows. Sec.2 is devoted to description of an individual element in our model. In Sec.3 we describe our model of the stream of coupled volumes. Numerical studies and results are presented in Sec.4. At last, Sec.5 concludes the paper with a general discussion.

## 2. Properties of individual electronic volume

Before the description of the model will be given, let's call attention to the dynamics of individual electronic pattern. Under electronic pattern we mean a confined volume of active medium, which consists of electron oscillators. The volume size is small in comparison with wavelength of radiation. It is supposed that the latter has cooperative or superradiative character [4,5,6]. This radiation arises as a result of electron interaction via the field of their radiation and phase focusing. Quite apparently that the similar processes can occur in the systems of nonlinear, nonisochronous oscillators only.

In this case the behaviour of M interacting electron oscillators is described by the following equations:

$$\dot{c}_k + F(c_k) = -\bar{c}, \tag{1}$$

where $c_k$ is the position of the k-th oscillator in the complex plane (k=1,...,M); $F(c_k)$ is a nonlinear function given by

$$F(c_k) = j\theta(|c_k|^2 - 1)c_k. \tag{2}$$

The mean field $\bar{c}$ is defined by $\bar{c} = \frac{1}{M}\sum_{k=1}^{M} c_k$ and specifies the dipole moment of electron ensemble. It is also a very useful macroscopic indicator of the system microscopic behaviour. Parameter θ is proportional to the ratio of the coefficient of nonlinearity and the electron dissipation coefficient [4].

Eqs.(1) were solved numerically with initial conditions

$$c_k(0) = \exp(-j(\varphi + \delta\cos\varphi)), \qquad (3)$$

where $\varphi=2\pi k/M$, $k=1,...,M$. The last expression shows the uniform electron distribution by phases in the interval $[0;2\pi]$ when small disturbance $\delta$ is presented.

The computations were carried out using the forth-order Runge-Kutta method with a timestep of 0.02. Fig.1 shows the square of absolute value $|\bar{c}|^2$ as a function of time for two different values of parameter $\theta$. It easy to see that there is a specific time T, after which electron's energy is lost and all of oscillations are dumped out. Hereafter we shall use this essential feature to construct our model.

## 3. The stream of coupled volumes: model's description

Let us consider the sequence of mutually coupled moving electronic patterns. In dimensionless complex variables the stream is governed by following differential equations:

$$\dot{c}_{ki} + F(c_{ki}) = -\bar{c}_i + K_F\bar{c}_{i-1} + K_B\bar{c}_{i+1}, \qquad (4)$$

$$k=1,...,M;\ i=1,...,N$$

Here N is the number of patterns existing in the flow simultaneously; $c_{ki}$ is dimensionless complex variable corresponding to field of the k-th electron in the i-th pattern; $K_F$ and $K_B$ are forward and backward coupling coefficients. The function $F(c_{ki})$ is the same as in Eq.(2). It is easily seen that Eq.(4) without two last terms describes the processes in each elementary pattern, discussed in Section 2. The mechanism of interactions in the flow is illustrated schematically at Fig.2.

Eqs.(4) were solved with initial conditions (3), the values of parameters $\theta$, $\delta$ and M are retained.

To simulate the dynamics of mutually coupled electronic patterns it is necessary to take into account their finite "lifetime" in the field of interaction. Referring back to Fig.1, we can see that at t exceeding the time T of radiation from elementary electronic volume, oscillations do not cease to exist. It is obviously that the "lifetime" should be close to T. The movement of electronic patterns is involved in numerical algorithm and fits the "transfer of information" from i-th to (i+1)-th stream section at regular intervals $\Delta t=T/N$, where N is the total number of patterns which exist in the stream simultaneously. Let us call $\Delta t$ "time of local interactions". Variable i indicates a fixed section of interaction field. It serves as a discrete analogue of spatial coordinate.

## 4. Numerical studies

As it evident from the foregoing, the model has quite a number of parameters. Let some of them be unchanged, namely $\theta$ and $\delta$. From Section 2 we know that these parameters have an essential impact on the shape of function $|\bar{c}|^2(t)$ and on the time T. Under this it is meant that the time of local interactions $\Delta t$ is limited. From the cases in Fig.3 one may conclude that finite limits of changing of $\Delta t$ exist. Outside these the oscillations in the stream are not possible: as $\Delta t$ increases, the time profile of $|\bar{c}|^2(t,i)$ simplifies. For this reason we assume $\Delta t$ to be fixed and equal to 1.

Under this we mean that the patterns arrive at the field of interaction at regular intervals. This is a simplified supposition; in real situation the intervals may be different and distinctions between them, as a rule, are devoid of regular character.

It can be shown that taking into account this fact leaves the dynamics of the stream qualitatively unchanged. Let $\Delta t=1+\xi$, where $\xi$ is random variable uniformly distributed on the interval $[-10^{-3};10^{-3}]$. A comparison of Poincare maps for the different sections of the stream is shown in Fig.4, 5 for $\Delta t=1$ (Fig.4) and $\Delta t=1+\xi$ (Fig.5). As it can be seen from Fig.5, the inclusion of small random variable $\xi$ (corresponding to weakly random modulation at the stream input) leads only to arising of noise component without qualitative changes of attractor. Moreover, identification of oscillatory regimes is hampered by noise. Hence from here on $\xi$ will be neglected.

Referring to Fig.4, as the spatial coordinate i increases, the dynamics of oscillations in the stream becomes more complicated. This fact is confirmed by power spectra shown in Fig.6.

The changing of $K_F$ and $K_B$ has an important bearing on the processes in the stream. Increasing of coupling coefficients leads to more complex spatiotemporal behaviour as it is illustrated by Fig.7 a, b.

All the above results have been obtained at N=10. What will be if the number of patterns presenting in the stream are simultaneously increased? Increase of N means decrease of local interactions' time $\Delta t$ and growth of frequency of electron entering into the stream. This effect is due to fixed time T of radiation from elementary electronic volume. When the number of patterns increases to N=150, the qualitative change in the spatiotemporal distribution of $|\bar{c}|^2$ was discovered. After a short transient process (initial spatial distribution of $|\bar{c}|^2$ was uniform) stationary in time and spatially irregular structures were observed. Fig.8 shows the evolution of spatial distribution profile as the coupling coefficients are increased. The spatial amplitude spectra for this case are shown in Fig.9.

We emphasize that the system is completely deterministic and clear of random influences. The varying of spectrum and profile of oscillations as $K_F$ and $K_B$ increases may denote arising of stationary spatial chaotic structures.

Some results of study of system behaviour under changes of parameters are stated also in [7,8].

## 5. Discussion and conclusions

In this paper we have presented and studied a model of pattern interactions in nonequilibrium dissipative media represented by the stream of coupled electronic volumes. The system is complex and has a very rich dynamics. It is worth of mentioning that each pattern in the stream is a globally coupled system, namely, it has a great number of degrees of freedom. The features of lumped system reflect individual dynamics of autooscillators, corresponding electron oscillators.

Now we would like to explain qualitatively the evolution in the stream of 10 coupled patterns. At the origin of the stream at i=2 there are two independed frequencies $f_1$ and $f_2$ in power spectrum (Fig.6, 7 b). In Poincare map we have 2-dimensional torus corresponding to quasiperiodicity. As the coordinate i increases, the new frequencies $f = mf_1 + nf_2$ appeare in the spectrum. The circle in Poincare map loose the smooth structure. This corresponds to the mixing of trajectories in the phase space. Further increasing of spatial coordinate in the stream leads to turbulent evolution of the system at $i=8\div10$. This is more visible when the coupling coefficients are large (Fig.7).

Let us compare the evolution described in this work with well-known scenario for lumped systems. If the spatial coordinate may be considered as control parameter in lumped system (with number of degrees of freedom, sufficient for chaos appearance), then observed transition to turbulence may occure in analogy to transient to chaos from quasiperiodicity in lumped system [9,10].

As for dynamics of the long stream involving a great number of electronic patterns, we consider it to be another feature of model versatility. Under this we mean that the quantitative transformation as N increasing leads to qualitative transition of system behaviour from spatialtemporal complicated oscillations to spatial chaos or irregular spatial patterns.

## References.

[1] Nonlinear waves. Structures and bifurcations. M.: Nauka, 1987.

[2] Kyhl R.L., Webster H.F. IRE Trans., 1956. ED-3. No4. pp.172-183.
[3] Ampilogova V.R., Zborovskii A.V., Trubetskov D.I., Hudzik K.V. / Lectures on Microwave Electronics and Radiophysics. Saratov. 1986. V.1. pp.106-110.
[4] Vainshtein L.A., Kleev A.I. / Cooperative Radiation of Electron Oscillators. DAN. 1990. V.311. No.4. pp.862-866.
[5] Ginzburg N.S., Sergeyev A.S. Superradiance in layers of excited classical and quantum oscillators. Zh. Eksp. Teor. Fiz. 1991. V.99. No.2. pp.438-446.
[6] Ginzburg N.S., Novozhilova Yu.V., Sergeev A.S. Superradiance of ensembles of classical electron-oscillators as a method for generation of ultrashort electromagnetic pulses. Nuclear Instruments and Methods in Physic Research A 341 (1994) pp.230-233.
[7] Mchedlova E.S., Trubetskov D.I. Radiation from a stream of small interacting electron oscillators. Technical Physics Letters. 1993. V.19. No.12. pp.784-786.
[8] Mchedlova E.S., Trubetskov D.I. Characteristics of radiation in chains of coupled small volumes containing electron oscillators. Technical Physics. 1994. V.39. No.10. pp.1061-1065.
[9] Rand D. and others. Universal transition erom quasiperiodicity to chaos in dissipative systems. Phys. Rev. Lett. 1982. V.49. N0.2. pp.132-135.
[10] Eckmann J.-P. Roads to turbulence in dissipative dynamical systems. Reviews of Modern Physics. V.53. No.4. Part I. 1981. pp.643-654.

**List of captions**

**Fig.1.** Dimensionless power of radiation in dependence on time for two different values of parameter $\theta$. The solutions are given by Eq.(1).

**Fig.2.** A schematic sketch of interactions in the stream.

**Fig.3.** Spatio-temporal distribution of the radiation field for different values $\Delta t$.

**Fig.4.** Sequence of Poincare sections for the system, described by Eq.(4); $\theta$=2, $\delta$=0.2, $K_F = 0.5$, $K_B = 0.5$, $\Delta t$=1, M=12; N=10. Each plot contains 6500 points; transient process equal to 500 points. Time step in numerical method was h=0.02.

**Fig.5.** Same as Fig.4, except that the stream is randomly modulated at the input, i.e. $\Delta t = 1 + \xi$, where $\xi$ is small random variable, $\xi \in [-10^{-3};10^{-3}]$.

**Fig.6.** Power spectra corresponding to Fig.4.

**Fig.7.** Sequence of Poincare sections (a) and power spectra (b) for time seria, given by Eq.(4), with the same parameters as in Fig.4, except that $K_F = K_B = 0.7$.

**Fig.8.** Changing of the shape of stationary spatial distribution for the long stream (N=150); $\theta$=4, $\delta$=0.2, M=12, $\Delta t$=0.02.

**Fig.9.** Spatial amplitude spectra corresponding to Fig.8.

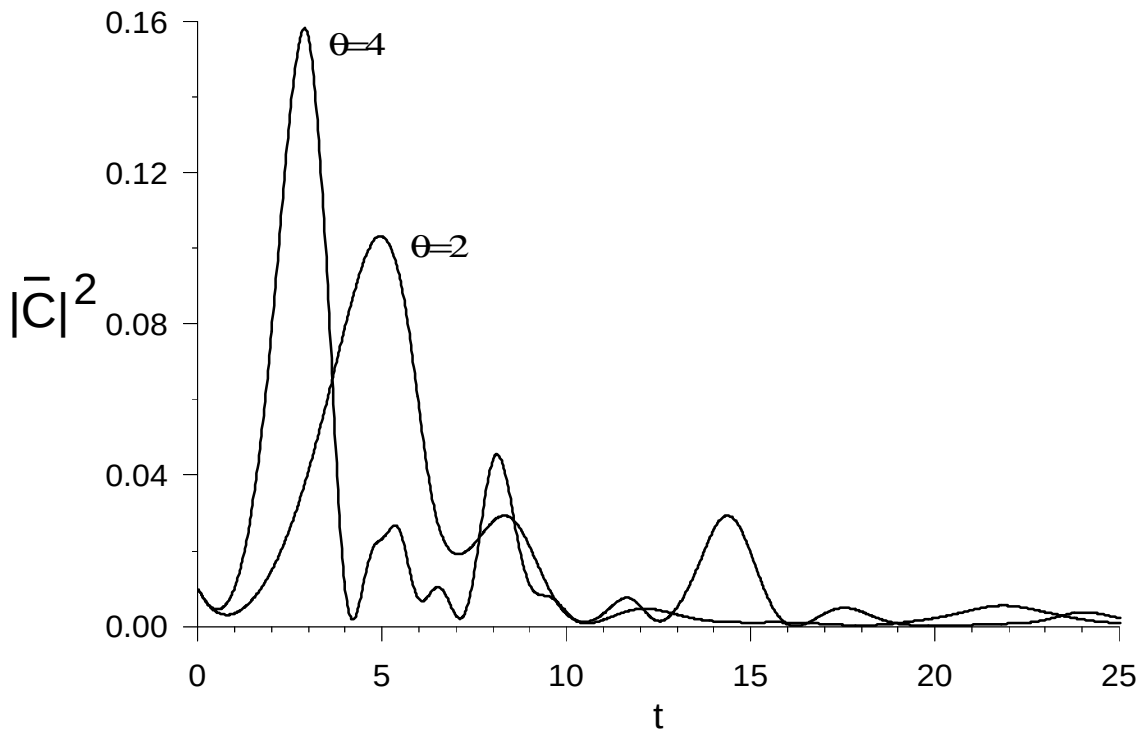

Fig.1.

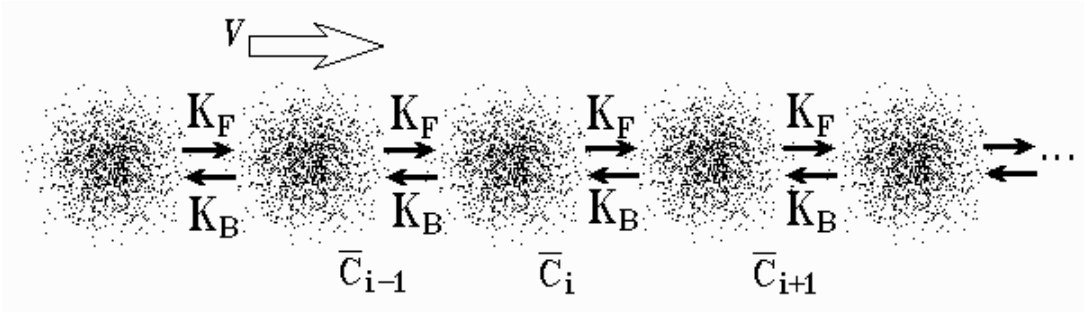

Fig.2.

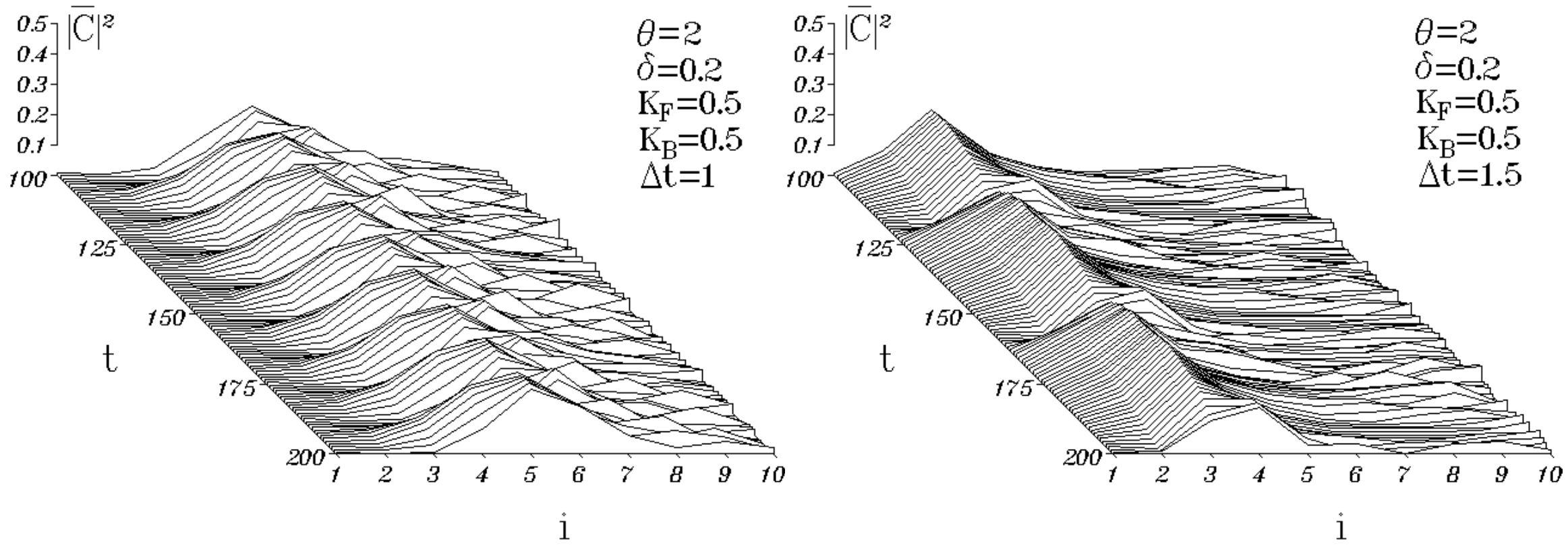

Fig.3.

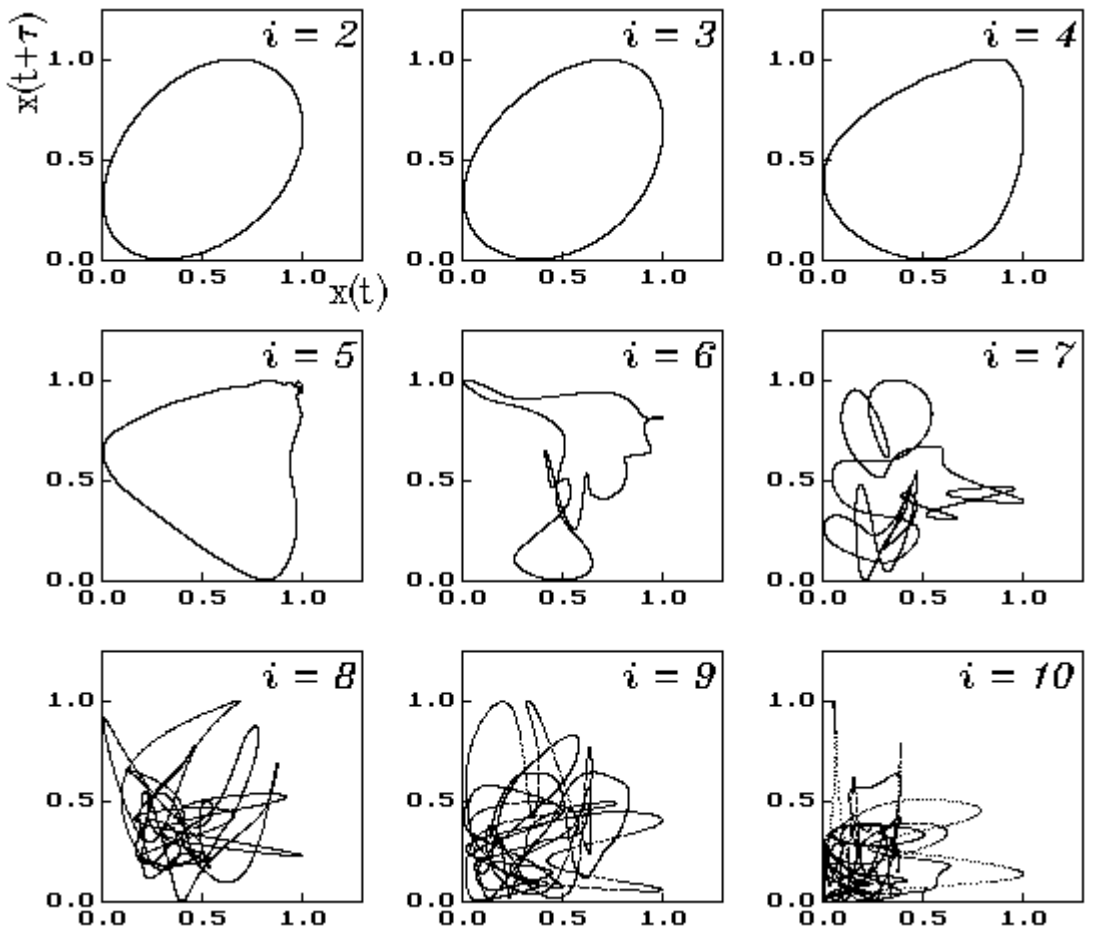

Fig.4

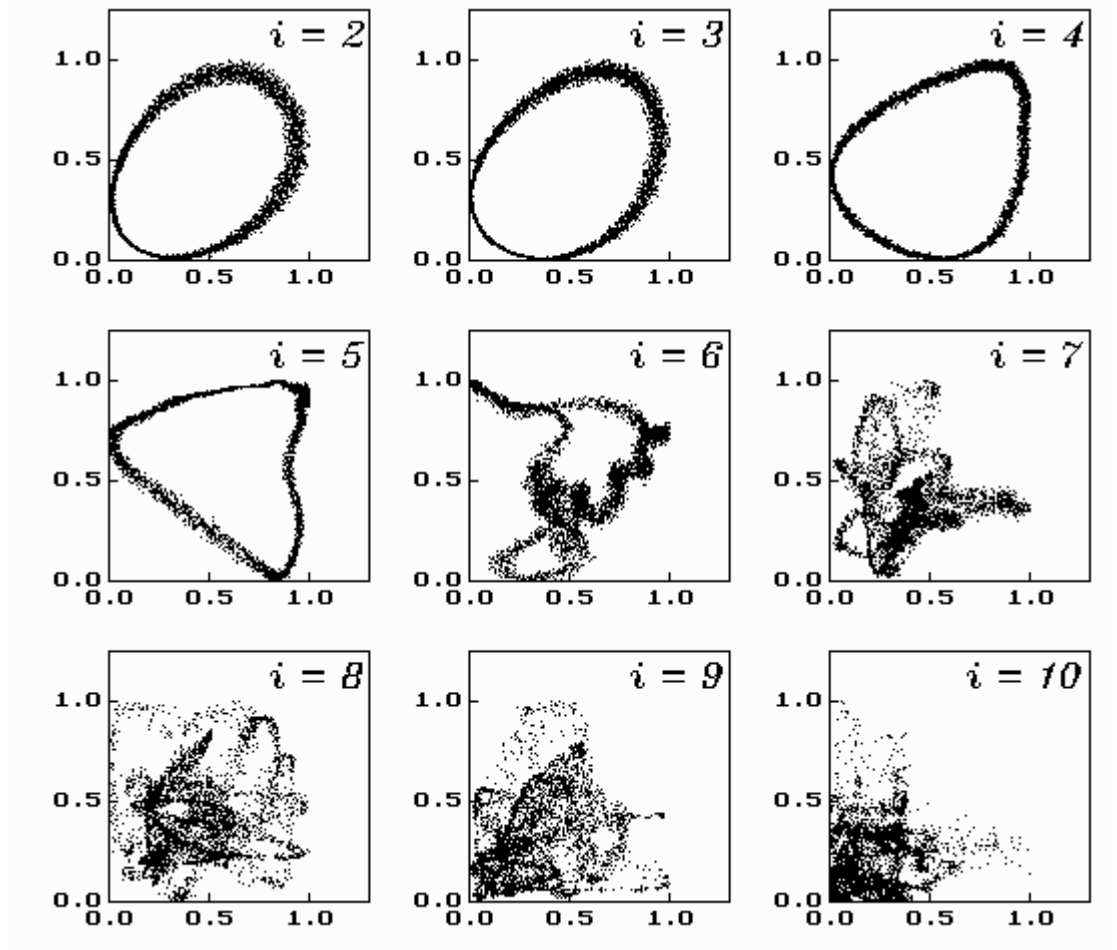

Fig.5

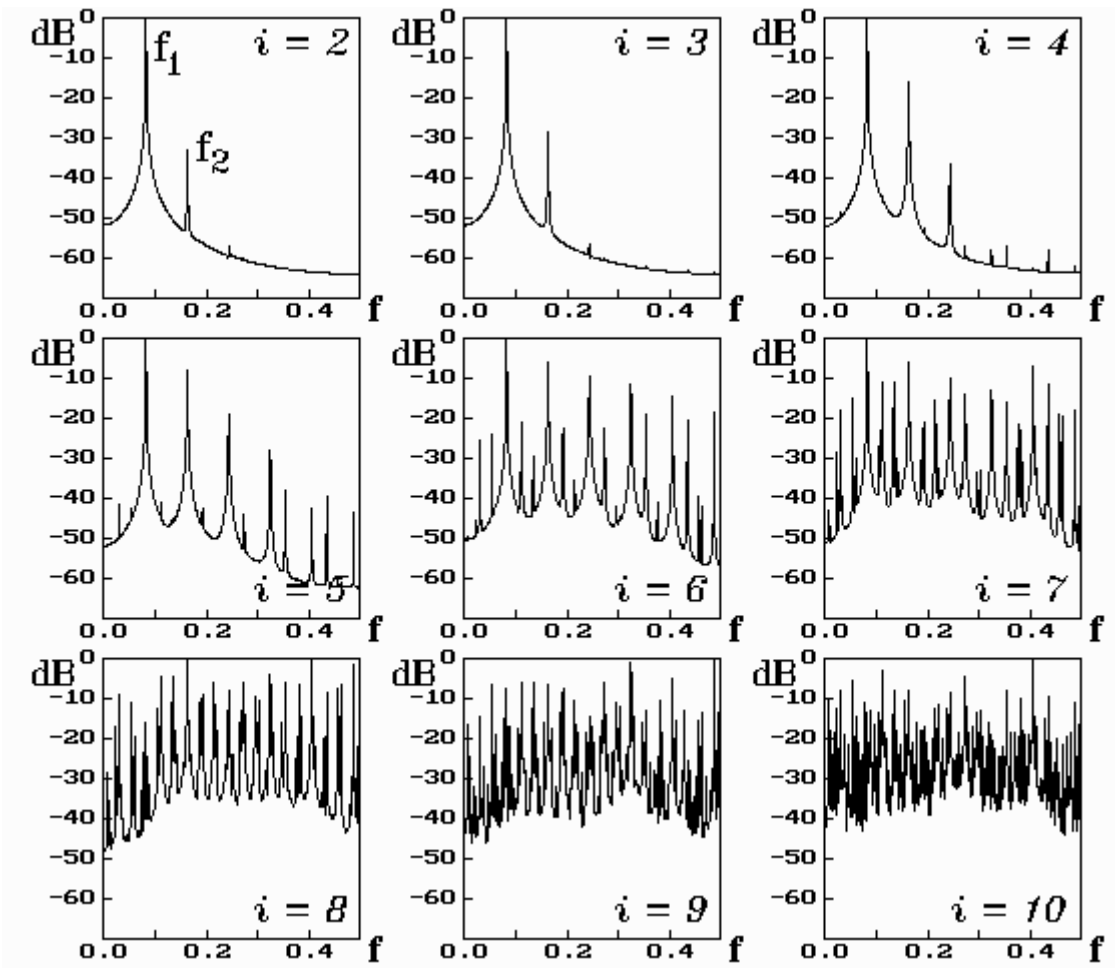

Fig.6

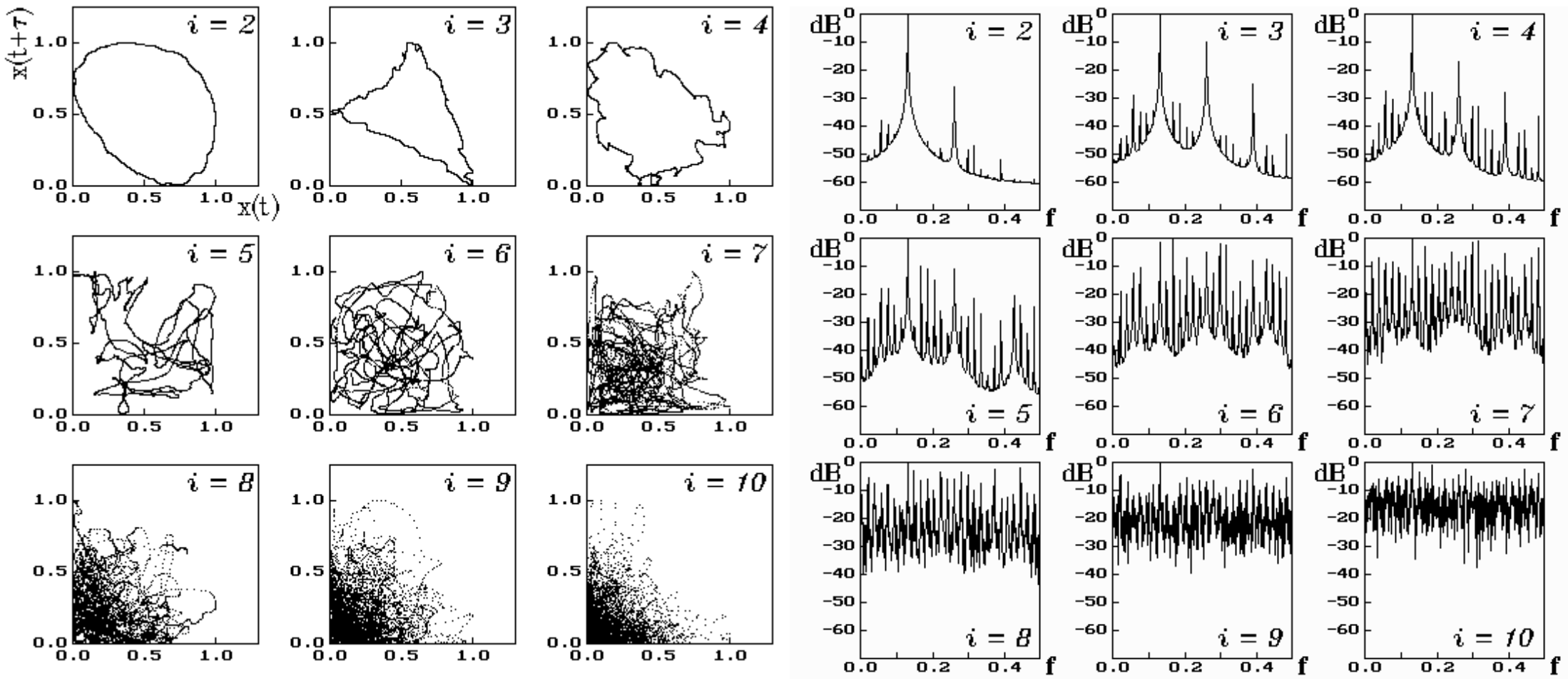

Fig.7 a

Fig.7 b

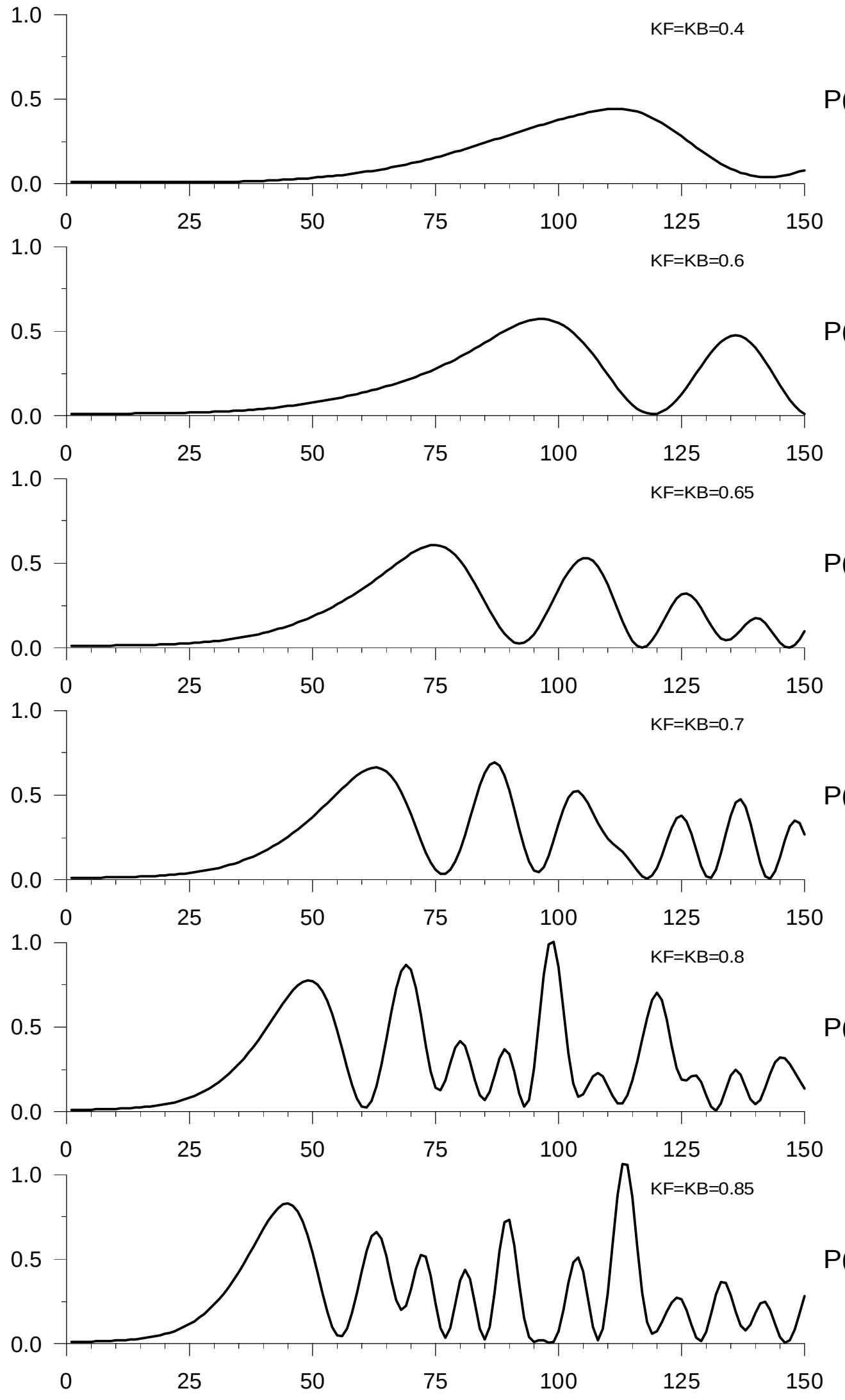

Fig.8

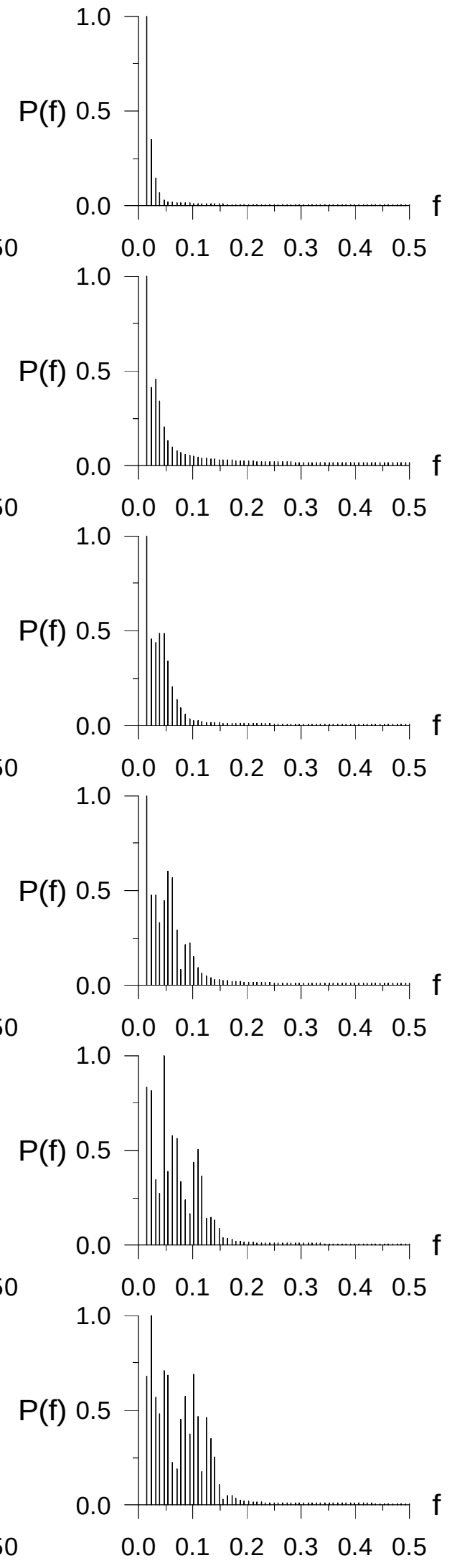

Fig.9